\documentclass[useAMS,usenatbib]{mn2e}
\usepackage[T1]{fontenc}
\usepackage{aecompl}
\usepackage{graphicx}
\usepackage{natbib}

\usepackage{ulem} 

\usepackage{times}
\usepackage{booktabs}
\usepackage{subfig}
\usepackage{color}
\usepackage{amsmath}

\def\lsim{\mathrel{\rlap{\lower 3pt\hbox{$\sim$}}\raise 2.0pt\hbox{$<$}}}
\def\gsim{\mathrel{\rlap{\lower 3pt\hbox{$\sim$}} \raise 2.0pt\hbox{$>$}}}

\def\aap{A\&A}

\def\apjl{ApJ}

\def\apjs{ApJS}

\title[PopIII stars and PopII/I GRBs]{PopIII signatures in the spectra of PopII/I GRBs}

\author[Ma et al.]{Q. Ma$^1$, U.~Maio$^{2,3}$, B.~Ciardi$^1$, R.~Salvaterra$^4$\\
$^1$ Max-Planck-Institut f\"ur Astrophysik, Karl-Schwarzschild-Stra\ss e 1, D-85748 Garching bei M\"unchen, Germany\\
$^2$ INAF -- Osservatorio Astronomico di Trieste, via G. Tiepolo 11, I-34131 Trieste, Italy \\
$^3$ Leibniz-Institut f\"ur Astrophysik, An der Sternwarte 16, D-14482 Potsdam, Germany\\
$^4$ INAF, IASF Milano, via E. Bassini 15, I-20133 Milano, Italy \\
}

\begin{document}

\date{Accepted?? Received??; in original form??}


\pubyear{2015}

\maketitle

\label{firstpage}

\begin{abstract}
We investigate signatures of population III (PopIII) stars in the metal-enriched environment of GRBs originating from population II-I (PopII/I) stars by using abundance ratios derived from numerical simulations that follow stellar evolution and chemical enrichment.   
We find that at $z>10$ more than $10\%$ of PopII/I GRBs explode in a medium previously enriched by PopIII stars (we refer to them as GRBII$\rightarrow$III).
Although the formation of GRBII$\rightarrow$III is more frequent than that of pristine PopIII GRBs (GRBIIIs), we find that the expected GRBII$\rightarrow$III observed rate is comparable to that of GRBIIIs, due to the usually larger luminosities of these latter.
GRBII$\rightarrow$III events take place preferentially in small proto-galaxies with stellar masses $\rm M_\star \sim 10^{4.5} - 10^7\,\rm M_\odot$, star formation rates $\rm SFR \sim 10^{-3}-10^{-1}\,\rm M_\odot/yr$ and metallicities $Z \sim 10^{-4}-10^{-2}\,\rm Z_\odot$.
On the other hand, galaxies with $Z < 10^{-2.8}\,\rm Z_\odot$ are dominated by metal enrichment from PopIII stars and should preferentially host GRBII$\rightarrow$III. 
Hence, measured GRB metal content below this limit could represent a strong evidence of enrichment by pristine stellar populations.
We discuss how to discriminate PopIII metal enrichment on the basis of various abundance ratios observable in the spectra of GRBs' afterglows.
By employing such analysis, we conclude that the currently known candidates at redshift $z\simeq 6$ -- i.e. GRB~050904 \cite[][]{2006Natur.440..184K} and GRB~130606A \cite[][]{2013arXiv1312.5631C} -- are likely not originated in environments pre-enriched by PopIII stars.
Abundance measurements for GRBs at $z\simeq 5$ -- such as GRB~100219A \cite[][]{2013MNRAS.428.3590T} and GRB~111008A \cite[][]{2014ApJ...785..150S} -- are still poor to draw definitive conclusions, although their hosts seem to be dominated by PopII/I pollution and do not show evident signatures of massive PopIII pre-enrichment.
\end{abstract}


\section{Introduction}
\label{sec:intro}
Primordial epochs host the formation of first stars and galaxies, and the consequent production of heavy elements in the Universe.
The basic understanding of the topic relies on cosmic molecule formation and gas cooling in growing dark-matter potential wells.
Features of primordial structures are still debated, though, and their observational signatures under investigation \cite[see e.g.][for recent reviews]{Ciardi2005, BrommYoshida2011}.
Early star formation episodes are expected to occur at redshift ($z$) larger than $\sim 10$, when the Universe was only half a Gyr old.
In these ages, star formation is led by pristine gas collapse that drives the birth of the so-called population III (PopIII) stars.
PopIII stars evolve and die ejecting the heavy elements they have synthesized in their cores.
These events pollute the surrounding medium and change its chemical composition.
Thus, the following generations of population II-I (PopII/I) stars will be born in a medium that has been pre-enriched of metals.
The transition from the primordial PopIII to the subsequent PopII/I regime is important to understand early cosmic phases, although the detailed properties of different populations are still a limiting unknown \cite[e.g.][]{2003Natur.425..812B,Schneider_etal_2003,Tornatore.Ferrara.Schneider_2007,Maio_etal_2010,2014arXiv1407.0034J,2014arXiv1405.7385Y}. 
For instance, PopIII stars have been predicted to have both very large and small masses \cite[e.g.][]{2002ApJ...571...30S,2011Sci...331.1040C,2014ApJ...785...73S}, while PopII/I stars have theoretical metal yields that often are affected by stellar evolution modeling.
The cosmological transition from one regime to the other is expected to take place at very early times, with a PopIII contribution to the cosmic star formation rate (SFR) dropping dramatically below $z\sim 10$ \cite[e.g.][]{Maio_etal_2010, Wise2012}.
The very first star formation sites are supposed to have dark-matter masses as small as $\sim 2\times 10^6\,\rm M_\odot$, to be extremely bursty and to feature specific SFRs up to $\sim 10-100\,\rm Gyr^{-1}$ \cite[][]{2013MNRAS.429.2718S, Biffi2013, deSouza2014, Wise2014}.
Such objects are enriched to average metallicities of $\sim 10^{-2}\, Z_\odot$ by $z > 9$.
\\
Direct observations of primordial times are quite problematic, because early (proto-)galaxies are usually faint and difficult to detect.
\\
Chemical imprints of PopIII stars may be retrieved from metal-poor stars of the Galactic halo \cite[][]{2008ApJ...679....6K,2009MNRAS.398.1782R} and low-mass halos at high redshift hosting damped Ly-alpha absorbers (DLAs)\cite[][]{2009ApJ...700.1672T,2013ApJ...772...93K, 2013MNRAS.435.1443M, 2014ApJ...787...64K}, but their features are still largely debated \cite[][]{2004ApJ...612..602T,2004ApJ...617..693D,2006A&A...451...19E,2007ApJ...665.1361T, 2011ApJ...730L..14K, 2010ApJ...712..435T,2010ApJ...717..542K}.
\\
A powerful tool to study this epoch is represented by gamma-ray bursts (GRBs), true {\it cosmic lighthouses} that have been observed up to $z\sim 8-9$ 
\cite[see e.g.][]{Salvaterra2009,Tanvir2009,Cucchiara2011} and can be employed for several purposes.
In particular, high-redshift GRBs can provide fundamental information about the early stages of structure formation and the properties of their own hosting galaxies, such as \cite[see ][for a more complete list]{McQuinn2009,Amati2013}:
metallicity and dust content
\cite[][]{Savaglio2005,Nuza2007,2011MNRAS.416.2760C, Mannucci2010, Elliott2014};
neutral-hydrogen fraction 
\cite[][]{Gallerani2008,Nagamine2008, McQuinn2008, RobertsonEllis2012};
local inter-galactic radiation field 
\cite[][]{Inoue2010};
stellar populations 
\cite[][]{2011MNRAS.416.2760C,Toma2011,deSouza2011,2013MNRAS.429.2718S,Wang2012,MaioBarkov2014};
early cosmic magnetic fields 
\cite[][]{Takahashi2011};
dark-matter models 
\cite[][]{deSouza2013, MaioViel2014arXiv};
and primordial non-Gaussianities 
\cite[][]{Maio2012nonG}.
\\
In the following, we will focus on the metal content of first GRB host sites in order to explore their gas chemical ratios and to constrain the corresponding stellar populations.
This is interesting for observers that try to infer the original stellar population from the species detected in the spectra of GRB afterglows.
This study will be performed by relying on published N-body hydrodynamical chemistry simulations including atomic and molecular non-equilibrium cooling, star formation, feedback mechanisms, stellar evolution, metal spreading according to yields and lifetimes of different populations (PopIII and PopII/I) for several chemical species \cite[][]{Maio_etal_2007, Maio_etal_2010, 2011MNRAS.416.2760C, 2013MNRAS.429.2718S}.
As the time evolution of the individual heavy elements is consistently followed within the simulation run, it will be possible to properly compute expected chemical ratios for PopIII and PopII/I GRB progenitors and to assess PopII/I GRBs exploding in environments possibly pre-enriched by former PopIII events.
\\
Throughout this work a standard $\Lambda$CDM cosmological model with total matter density parameter $\Omega_{0,m}=0.3$, dark energy density parameter $\Omega_{0,\Lambda}=0.7$, baryonic matter density parameter $\Omega_{0,b}=0.04$, Hubble constant in units of 100~km~s$^{-1}$~Mpc$^{-1}$ $h=0.7$, spectral normalization $\sigma_8=0.9$ and primordial spectral index $n=1$ is assumed.
\\
The paper is organized as follows:
in Section~\ref{sec:sel} we describe the simulations employed in this work; 
in Section~\ref{sec:res} we discuss the results in term of rate of GRBs, metal signatures and properties of the host galaxies;
in Section~\ref{sec:conclusion} we give our Conclusions.

\section{Simulations and metal abundances}
\label{sec:sel}
In this paper we use the simulations presented in \cite{Maio_etal_2010}, of which we  describe the essential features, while we refer the reader to the original paper for more details. The same simulations have already been used in \cite{2011MNRAS.416.2760C} and \cite{2013MNRAS.429.2718S} to study the properties of high-$z$ GRBs and their host galaxies.
In the following, we will refer to the previous papers as C2011 and S2013, respectively.
In the N-body hydrodynamical chemistry calculations, the transition from a primordial (PopIII) to a PopII/I star formation regime is determined by the metallicity, $Z$, of the star forming gas. More specifically, a Salpeter initial mass function (IMF; $ \phi (m_\star) \propto m_\star^{-2.35}$ is the number fraction of stars per unit stellar mass) with mass range $\rm [100 - 500]~M_{\odot}$ is adopted if $Z<Z_{crit}=10^{-4}Z_{\odot}$, while a Salpeter IMF with mass range $\rm [0.1 - 100]~M_{\odot}$ is used at larger metallicities. 
\\
Besides gravity and hydrodynamics, the code follows early chemistry evolution, cooling, star formation and metal spreading from all the phases of stellar evolution according to the suited metal-dependent stellar yields (for He, C, N, O, S, Si, Mg, Fe, etc.) and mass-dependent stellar lifetimes for both the pristine PopIII regime and the metal-enriched PopII/I regime \cite[][]{TBDM2007, Maio_etal_2007, Maio_etal_2010}.
Yields for massive pair-instability supernovae (PISN) in the stellar mass range $\rm [140-260]\,M_\odot$ are taken by \cite{HegerWoosley2002}.
Standard type-II supernova (SNII) yields for progenitors with different metallicities and masses of $\rm [8-40]\,M_\odot$ are from \cite{WW1995}, AGB metal production for lower-mass stars is followed according to \cite{1997A&AS..123..305V} and type-Ia SN (SNIa) yields are from \cite{Thielemann_et_al_2003}.
Several models for stellar nucleosynthesis are available in the literature and can give different yields; however, the overall metallicity is usually affected in a minor way \cite[as already discussed in e.g.][]{Maio_etal_2010}.\\
In general, to better characterize the metal content of gas, stars or galaxies, it is convenient to define abundance ratios between two arbitrary species $A$ and $B$ as follows:
\begin{equation}
[{\rm A}/{\rm B}] = \textup{log$_{10}$}\left(\frac{N_{\rm A}}{N_{\rm B}}\right)-\textup{log$_{10}$}\left(\frac{N_{\rm A}}{N_{\rm B}}\right)_{\odot},
\label{ratios}
\end{equation}
where $ N_{\rm A} $ and $N_{\rm B}$ are the number densities of the two species and the subscript ${}_{\odot}$ refers to the corresponding solar content (Table \ref{sunme};
\citealt{Asplund2009}).
Abundance ratios are powerful tools to investigate the pollution history and to probe stellar evolution models against observational data.
\begin{table}
\centering
\begin{tabular}{l c c c}
\hline
X   & log$_{10}(N_{\rm X}/N_{\rm H})_{\odot}$  & X   &log$_{10}(N_{\rm X}/N_{\rm H})_{\odot}$\\
\hline
C    &-3.57             & Si   & -4.49 \\
N    & -4.17            & S    & -4.86 \\
O    & -3.31            & Ca   & -5.66 \\
Mg   & -4.4             & Fe   & -4.53 \\
\hline
\end{tabular}
\caption{Solar element abundance adopted in this paper for calculating the chemical element abundance (\citealt{Asplund2009}).}
\label{sunme}
\end{table}

\section{Results}
\label{sec:res}
In this Section we will show results on the rate of GRBs originated from PopII/I stars exploding in a gas enriched by PopIII stars, together with a discussion on the properties of the galaxies hosting such GRBs and observational strategies to identify them.
This is an interesting issue, because high-redshift PopII/I GRBs exploding in a medium pre-enriched by former star formation episodes can reveal features of primordial PopIII generations through the metallicity patterns of their hosting galaxy.

\subsection{Rate of GRBs}
\label{rate}

We calculate the rate of GRBs as described in C2011. Here we provide only the essential information, while we refer the reader to the original paper for more details.
The comoving GRB formation rate density, $\rho_{GRBi}$, can be calculated as:
\begin{equation}
  \rho_{GRBi}=f_{GRBi}\zeta_{BH,i}\rho_{\ast,i},
  \label{grb}
\end{equation}
where $i$ indicates the subsample of GRBs considered, $f_{GRBi}$ is the fraction of BHs that can produce a GRB, $\zeta_{BH,i}$ is the fraction of BHs formed per unit stellar mass and $\rho_{\ast,i}$ is the comoving star formation rate density. We select star forming particles associated to PopII/I stars based on the criterium that $Z>Z_{crit}$, i.e. without an upper metallicity cut, as the GRB1 in C2011.
For the sake of clarity though, here we change the nomenclature and we refer to GRBII (GRBIII) as those GRBs originating from PopII/I (PopIII) stars.
These are called GRB1 (GRB3) in C2011.
Following the original paper, we adopt $\zeta_{BH,II}=0.002$ and $f_{GRBII}=0.028$ for GRBII (which correspond to a minimum stellar mass of stars dying as BHs of 20~M$_\odot$), and $\zeta_{BH,III}=0.0032$ for GRBIII. $f_{GRBIII_{up2}}=0.0116$ is obtained assuming that no GRBIII is present among the 266 GRBs with measured redshift, while the more stringent value $f_{GRBIII_{up1}}=0.0039$ is found imposing that none of the 773 GRBs observed by {\tt Swift} is a GRBIII.
We additionally study a third class of GRBs, namely the GRBII that explode in a medium enriched by PopIII stars, which we indicate as GRBII$\rightarrow$III. Here, $\zeta_{BH,II\rightarrow III}=\zeta_{BH,II}$, $f_{GRBII\rightarrow III}=f_{GRBII}$, and $\rho_{\ast,II\rightarrow III}$ is explicitly calculated by selecting only those star forming particles which have been enriched by PopIII stars.
This task is accomplished by using indicative metal ratios (see also next) for PopIII enrichment, consistently with the mentioned yield tables.
In particular, PopIII PISN enrichment results characterized by:
$\rm [Fe/O]<0$, 
$\rm [Si/O]>0$, 
$\rm [S/O]>0$ and 
$\rm [C/O]<-0.5$.
These limits are used as reference selection criteria for PopII/I star forming regions enriched by PopIII stars (see Sec.~\ref{sec:metals} for further discussion).
\\
\begin{figure}
  \centering
  \includegraphics[width=1.0\linewidth]{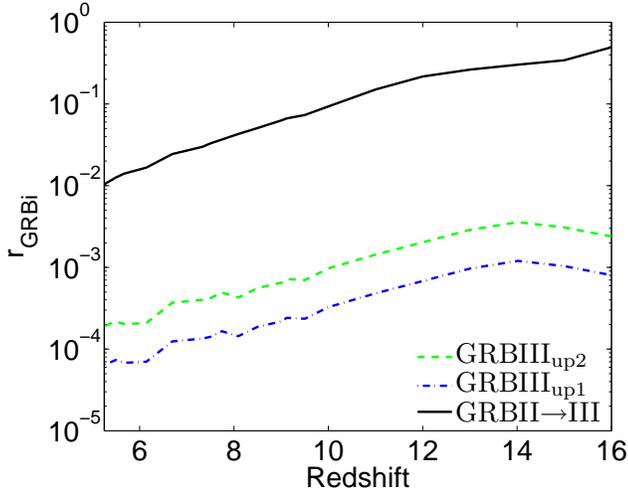}
  \caption{Redshift evolution of the ratio between the rate of GRB$i$ and the total GRB rate. $i$=II$\rightarrow$III (black solid) and III (blue dotted-dashed for GRBIII$_{up1}$ and green dashed for GRBIII$_{up2}$).
}
\label{ratio}
\end{figure}
Figure~\ref{ratio} shows the redshift evolution of the quantity $r_{GRBi}=\rho_{GRBi}/(\rho_{GRBtot})$, where $\rho_{GRBtot}$ is the total GRB rate.
As expected, both $r_{GRBIII}$ and $r_{GRBII\rightarrow III}$ decrease with decreasing redshift, as the number of PopIII stars and their impact on the gas metal enrichment become less relevant. As a consequence, while $r_{GRBII\rightarrow III}$ is as high as $\sim 50\%$ at $z\sim 16$, it becomes less than $1\%$ at $z<5$. On the other hand, $r_{GRBIII}$ is always negligible, consistently with previous studies, as the transition from PopIII to PopII/I stars is very rapid.\\
The (physical) observable rate of GRB$i$ (in units of yr$^{-1}$sr$^{-1}$) at redshift larger than $z$ can be written as (e.g. C2011):
\begin{eqnarray}
  \label{number}
  R_{GRBi}(>z) & = &      \gamma_b\int_z dz'\frac{dV(z')}{dz'} \frac{1}{4\pi} 
  \frac{\rho_{GRBi}(z')}{(1+z')} \nonumber \\
  & \times & \int_{L_{th}(z')} dL' \psi_i(L'),
  \label{eq:GRBrate}
\end{eqnarray}
where
$\gamma_b=5.5 \times 10^{-3}$ is the beaming fraction for an average jet opening angle of $\sim 6^\circ$ \cite[][]{Ghirlanda2007,Ghirlanda2013}, 
$dV(z)/dz$ is the comoving volume element and 
$\psi_i$ is the normalized GRB luminosity function.
The factor $(1 + z)^{-1}$ accounts for the cosmological time dilation.
The last integral gives the fraction of GRBs with isotropic equivalent peak luminosities above $L_{th}$, corresponding to an observed photon flux of $\sim 0.4$ ph s$^{-1}$ cm$^{-2}$ in the 15-150 keV band of the {\tt Swift/BAT}.
We refer the reader to C2011 for a more exhaustive discussion on the choice of $L_{th}$ and $\psi_i$ \cite[see also][]{Salvaterra2012}.
\\
Interestingly, while there are much less GRBIII than GRBII$\rightarrow$III (Fig.~\ref{ratio}), the former can be more easily detected because of their larger luminosity, as can be seen from Figure~\ref{rate1}, resulting in very similar observable rates.
Nevertheless, we should note that $R_{GRBIII}$ has to be considered as an upper limit to the expected number of detections, as we have assumed here (as in C2011) that all GRBIII can be detected by {\tt Swift/BAT} thanks to their expected extreme brightness.
This should be true for GRBs originating from massive PopIII stars (considered in this paper), while in case of more standard masses luminosities should be similar to those from regular PopII stars\footnote{
  GRBIII rates from C2011 have been updated by taking into account the new GRBs detected by {\tt Swift} after the publication of C2011. These lead to a decrease of the C2011's PopIII GRB rate by a factor of $773/500=1.546$ for the 'up1' case and by a factor of $266/140=1.9$ for the 'up2' case.
}
.

\begin{figure}
  \centering
  \includegraphics[width=1.0\linewidth]{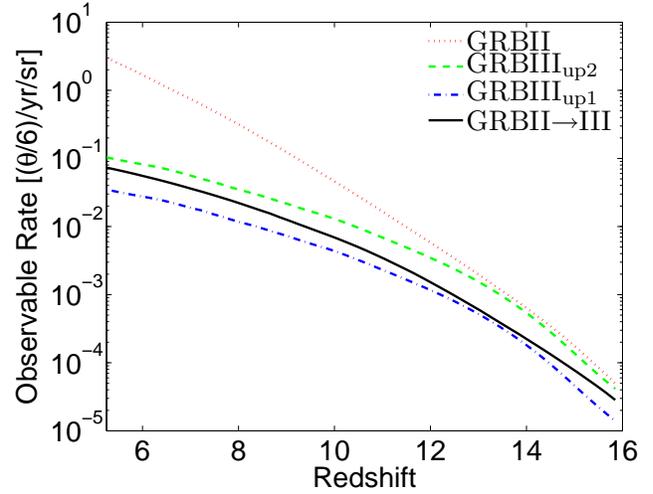}
  \caption{Rate of GRB$i$ with redshift larger than $z$ observable by {\tt Swift/BAT} (see text for details). $i$=II (red dotted line), II$\rightarrow$III (black solid) and III (blue dotted-dashed for GRBIII$_{up1}$ and green dashed for GRBIII$_{up2}$).
  }
  \label{rate1}
\end{figure}

\subsection{Metal signatures}
\label{sec:metals}

\begin{figure}
  \centering
      {\includegraphics[width=3in]{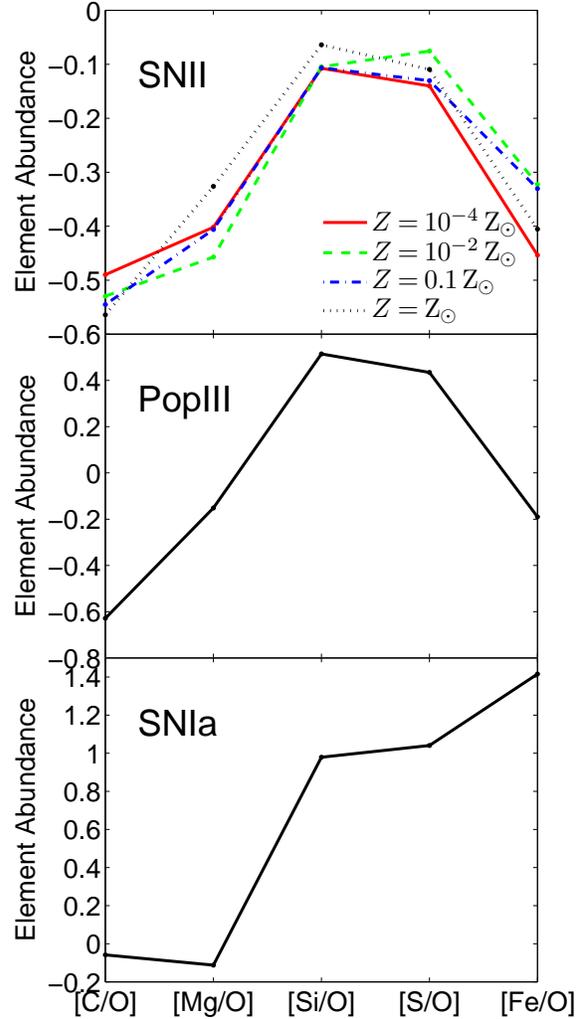}}
      \caption{Element abundance as produced by: SNII (upper panel), PISN stars (middle) and SNIa (lower). The contribution from SNII depends on the gas metallicity, $Z$, as indicated in the legend.
      }
      \label{mm1}
\end{figure}

\begin{figure}
\centerline
	{\includegraphics[width=2.6in]{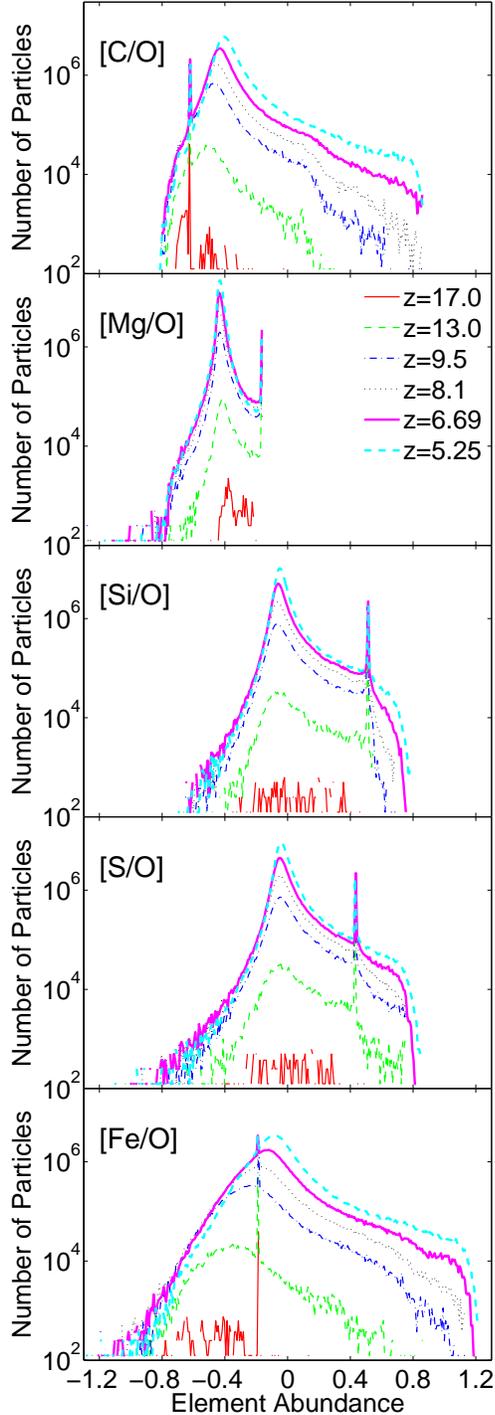}}   
	\caption{Abundance distribution (in number of particles) of, from top to bottom, C, Mg, Si, S and Fe of star forming particles with $Z>Z_{crit}$. The distributions are shown at $z=17$ (solid red lines), 13 (green), 9.5 (blue), 8.1 (black), 6.69 (magenta) and 5.25 (cyan). 
}
\label{odis}
\end{figure}

We now discuss the metal signatures which could help in identifying a GRBII$\rightarrow$III.
Since the contribution to the metal enrichment from different stellar types is explicitly included in the simulations, the final abundance of each heavy element can be computed from the metal masses as traced at each output time.
\\
We use abundances normalized to oxygen, because this species is not very sensitive to theoretical yield uncertainties and is widely adopted in the literature.
Additional reference species will be discussed in the following, though.
For a given species X, we calculate the abundance ratio [X/O] in eq.~\ref{ratios} with 
$N_{\rm X}/N_{\rm O} = \left( A_{\rm O}m_{\rm X} \right) / \left( A_{\rm X}m_{\rm O} \right)$, where 
$A_{\rm X}$ is the atomic mass number of element X and 
$m_{\rm X}$ is the mass of X in a particle as given by the simulation.
Moreover, it is possible to write
${\rm log}_{10} \left( N_{\rm X}/N_{\rm O} \right)_\odot = \left[ {\rm log}_{10} \left( N_{\rm X}/N_{\rm H} \right)_\odot- {\rm log}_{10} \left( N_{\rm O}/N_{\rm H} \right)_\odot \right]$.
Since GRBs are usually found in actively star forming environments, only gas particles with non-zero SFR are considered while performing the calculations of the various chemical abundances.
\\
Before discussing the element abundances obtained from the simulations, it is instructive to look at Figure \ref{mm1}, which shows the metal abundance of some elements from different stellar types of a simple stellar population, as expected by averaging over the corresponding IMF. More specifically, [X/O] in eq.~\ref{ratio} is obtained using 
$N_{\rm X}/N_{\rm O} = \left[ A_{\rm O} \int M_{\rm X}(m_\star) \phi (m_\star) dm_\star \right] / \left[ A_{\rm X} \int M_{\rm O}(m_\star) \phi (m_\star) dm_\star \right]$, where $M_{\rm X}$ is the stellar mass yield of element X for a given stellar population and the integration is performed over the relevant mass range for each stellar type.
In the Figure, we do not show AGB stars, as their contribution to the metal enrichment is negligible for all shown elements, with the exception of C and O, which are produced in different amounts according to progenitor masses and metallicities, although typically [C/O]$\gsim 0$.
\\
It is evident that each stellar phase is characterized by specific yields, that can be used to exclude or identify their contribution to metal enrichment. In particular, a positive and large value of [Fe/O] would indicate a strong contribution from SNIa, while a negative value could be due both to SNII and to PopIII stars (PISN). On the other hand, while SNII contribute with negative [Si/O] and [S/O], PopIII stars and SNIa contribute with positive ratios of the same elements.
As mentioned above, AGB stars are the only ones to contribute with a usually positive [C/O] ratio.
It is then clear that we can easily exclude gas enriched by SNIa as the one with [Fe/O]$>$0. Similarly, [S/O]$<$0 and [Si/O]$<$0 indicates gas enriched by SNII, and [C/O]$>$0 by AGB stars.
It should be noted that the gas metallicity, $Z$, can affect the outflow of SNII and of AGB stars\footnote{SNIa yields are weakly dependent on metallicities because they all derive from a white dwarf growing towards the Chandrasekhar mass and burning (in electron-degenerate gas) C and O in equal proportions \cite[][]{Thielemann2004}.}, but the considerations above still apply. So, to ensure that gas enriched by a stellar type other than PopIII stars is excluded, criteria such as  [Fe/O]$<$0, [S/O]$>$0, [Si/O]$>$0 and [C/O]$<$0, could be adopted, although, as mentioned previously, the more stringent condition [C/O]$<$-0.5 has been chosen (see Sec.~\ref{rate}).
\\
Figure~\ref{odis} shows the distribution at various redshifts of some relevant elements in star forming particles with $Z>Z_{crit}$.
With the exception of the highest redshift, when there are only a handful of star forming particles, two peaks can be clearly identified in each curve. The broader peak increases with decreasing redshift, and its  metallicity abundance is similar to the one from SNII, so that metal enrichment of the particles in this peak might be led by gas expelled during SNII explosions. The increase of the peak with $z$ is due to the time and metallicity evolution of the involved stellar systems. Indeed the relative shift in carbon abundance [C/O] is due to the metallicity-dependent carbon yields of massive stars augmented by the delayed release of carbon from low- and intermediate-mass stars \cite[e.g.][]{Akerman2004, Cescutti2009, Cooke2011}. Thus, the low values observed at $ z\sim 17-13$ are caused by SNII exploding in increasingly enriched gas (i.e. producing increasingly lower [C/O], as depicted in Fig.~\ref{mm1}) and a subsequent rise at $z\sim 13-5$ led by AGB stars.
The similar trend observed in [Fe/O] is a results of cosmic metal evolution as well, and at $z\sim 5-6$ the ratio is sustained also by the very first SNIa starting to explode after $\sim 1$~Gyr lifetime.
Conversely, there is little variation in the distribution of the $\alpha$ ratios ([Si/O], [S/O], [Mg/O]), as the nuclear reactions leading to their formation path are directly linked to O production.
The element abundance of the narrow peaks seems more consistent with the yields from PopIII stars. Below $z \sim 9$ the peak stops increasing as a result of the extremely small PopIII SFR.
\\
Because some metallicity limits are equivalent in excluding particles enriched by a specific stellar type (e.g. [Si/O], [S/O] and [Mg/O] can be equally used to exclude particles with metallicity dominated by SNII, while [C/O] or [Fe/O] can be used for SNIa), GRBII$\rightarrow$III can be identified also using less stringent criteria, and typically only two conditions are enough, such as [C/O] plus [Si/O], [C/O] plus [Mg/O], [Mg/O] plus [Fe/O] or [Si/O] plus [Fe/O].
We find that if we adopt [C/O]$<$-0.5 and [Si/O]$>$0, or [C/O]$<$-0.5 and [Mg/O]$>$-0.4 the ratio $r_{GRBII\rightarrow III}$ is $\sim 1\%$ (4\%) and $\sim 20\%$ (50\%) larger then the one shown in Figure~\ref{ratio} at $z=17$ (5).
At $z>12$ the following criteria give similar results: [C/O]$<$-0.5 and [Mg/O]$>$-0.4, [Fe/O]$<$-0.1 and [Si/O]$>$0, [Fe/O]$<$-0.1 and [Mg/O]$>$-0.4.
Finally, [S/O] and [Si/O] have similar selection effects, i.e. the same particles are selected if we use either [S/O]$>$0 or [Si/O]$>$0.

\subsection{Host galaxies}

In this Section we discuss some properties of galaxies hosting a GRBII$\rightarrow$III.
Both observations and theoretical studies suggest that high-redshift ($ z > 5$) long GRBs occur in objects with a SFR lower than the one of the local Universe ($ z \sim 0$).
\citet{2013MNRAS.429.2718S}, for example, find that such host galaxies have typically a stellar mass $M_{\star} \sim 10^{6}-10^8$~M$_{\odot} $ and a SFR $\sim 0.003-0.3$~M$_{\odot}$~yr$^{-1}$.
\begin{figure}
        \centering
        \includegraphics[width=1.0\linewidth]{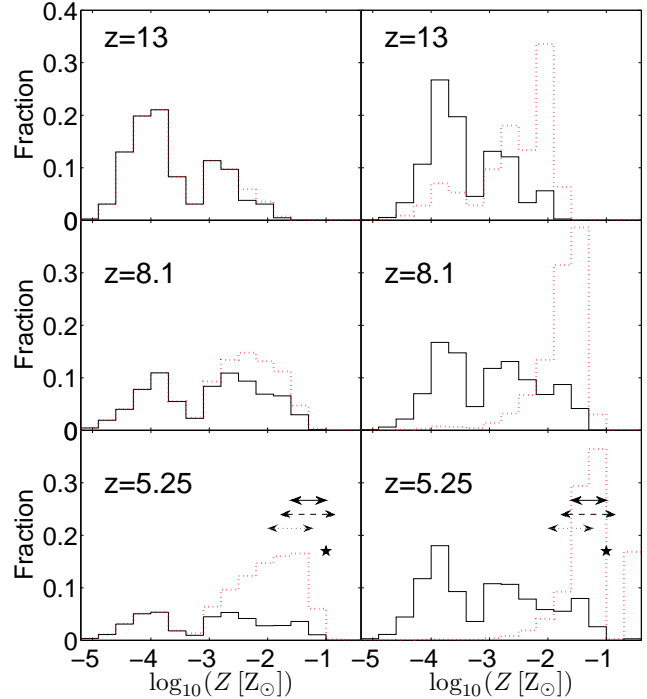}
        \caption{Metallicity distribution of halos hosting a GRBII$\rightarrow$III (solid black line) and of all PopII/I star forming halos (dotted red) at redshift $z=13$ (top), 8.1 (middle) and 5.25 (bottom). {\it Left panels:} halos number fraction distribution normalized to the total number of star forming halos. {\it Right panels:} the distributions are ``weighted'' by their respective SFRs and normalized to the total SFR (see text for more details). The double arrows denote the $Z$ range of GRB~050904 (solid black), GRB~130606A (dashed black) and GRB~111008A (dotted black). The black solid star denotes the $Z$ of GRB~100219A.
}
\label{fig:Z}
\end{figure}
In general, metal enrichment from primordial star forming regions is efficient in polluting the structures hosting GRB episodes and in bringing typical metallicities above the critical value for stellar population transition.
This is in agreement with recent theoretical studies of primordial galaxies \cite[][]{Maio2011, Biffi2013, Wise2012, Wise2014} and of GRB hosts \cite[][]{2011MNRAS.416.2760C, 2013MNRAS.429.2718S}, that suggest a typical metallicity range between $\sim 10^{-4}\,\rm Z_\odot$ and $\sim 10^{-1}\,\rm Z_\odot$ at $z > 6$.
However, it is not clear yet how these statistical trends change when considering the sample of GRBII episodes in regions polluted by previous PopIII events.
In order to discuss this issue, we compute the metallicity distributions at different redshift for both the whole galaxy population and for the GRBII$\rightarrow$III hosting population.
\\
In Figure~\ref{fig:Z} we show the metallicity distribution of halos hosting a GRBII$\rightarrow$III (solid black line) and of all PopII/I star forming halos (dotted red line) at different redshift. While at the highest redshift both distributions are very similar in terms of halo number fractions (left panels), differences are visible at later times, when the typical metallicity of GRBII hosts increases.
The simplest interpretation of these trends relies on the fact that primordial GRBII events take place in environments that have been previously enriched by the first PopIII generation and hence the two distributions almost coincide at early epochs, when only PopIII stars had time to explode and pollute the gas.
At later stages, PopIII stars play an increasingly minor role in the global star (and GRB) formation rate, which is dominated by PopII/I stars.
This means that successive GRBII events will happen in gas that has been enriched by several generations of PopII/I stars, as already visible in the high-$Z$ tail of the distributions at $z=8.1$.
Consequently, GRBII$\rightarrow$III can survive only in the low-$Z$ tail, where there is still some probability of having residual PopIII pollution in galaxies with weak star formation. This can be more clearly seen from the panels in the right columns, where we show a SFR ``weighted'' metallicity distribution, i.e. the fraction SFR$_{i,j}$/SFR$_{i,tot}$, 
where SFR$_{i,j}$ is the SFR of halos in the $j$-th metallicity bin calculated for the component $i=$II, II$\rightarrow$III, and SFR$_{i,tot}$ is the SFR of all halos for the same component.
This explains the distributions at $z=5.25$ and, more quantitatively, suggests a metallicity of roughly $Z<10^{-2.8} \,\rm ~Z_\odot$ to highlight the presence of a GRBII$\rightarrow$III (or possibly a GRBIII) in the early Universe.
Another distinctive feature is the presence of two peaks at $Z \sim 10^{-4}$~Z$_\odot$ and $10^{-2.8}$~Z$_\odot$, found also in C2011 for GRBIII host galaxies (see their Fig.~3).
This may be an indication that GRBII$\rightarrow$IIIs reside in galaxies similar to those hosting GRBIII, albeit with a higher star formation activity, as better highlighted by weighting the distribution of the host metallicities by the SFR (right panels).
Finally, we should note that a comparison with the metallicity range of the host galaxy of GRB~130606A \citep{2013arXiv1312.5631C,Chornock2013} seems to indicate that this is not a GRBII$\rightarrow$III. Although the redshift of GRB~130606A is 5.91, i.e. slightly higher than what shown in the Figure, this does not change our conclusion because of the small redshift evolution of the peak at $Z\sim 10^{-2}-10^{-1}$~Z$_\odot$.
A similar conclusion can be drawn for GRB~050904 at $z=6.3$ \citep{2006Natur.440..184K},  GRB~111008A at  $z=5.0$ \citep{2014ApJ...785..150S} and GRB~100219A at $z=4.7$ \citep{2013MNRAS.428.3590T}.
\\
\begin{figure}
  \centering
  \includegraphics[width=1.0\linewidth]{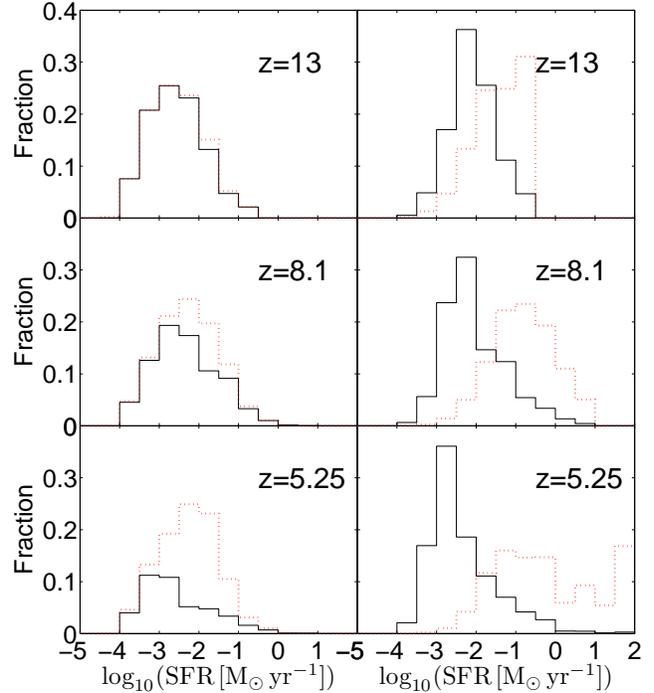}
  \caption{SFR distribution of halos hosting a GRBII$\rightarrow$III (solid black line) and of all PopII/I star forming halos (dotted red) at redshift $z=13$ (top), 8.1 (middle) and 5.25 (bottom). {\it Left panels:} halos number fraction distribution normalized to the total number of star forming halos. {\it Right panels:} the distributions are ``weighted'' by their respective SFRs and normalized to the total SFR (see text for more detail).
  }
  \label{fig:sfr}
\end{figure}
The corresponding SFR distributions, shown in Figure~\ref{fig:sfr}, span the range $\sim 10^{-4.5} - 10\,\rm  M_\odot$~yr$^{-1}$ and neatly support the interpretation, giving an upper limit of $\sim 10^{-2}\,\rm  M_\odot$~yr$^{-1}$ for GRBII$\rightarrow$III host galaxies (higher values of the SFR are found, but with a much lower probability).
More specifically, the SFR displays a trend similar to the one of the metallicity, with GRBII$\rightarrow$III forming preferentially in halos with a SFR smaller than the one of GRBII hosts, and a large fraction of halos with $\rm SFR \sim 10^{-4}-10^{-3}~M_{\odot}~yr^{-1} $ hosting a GRBII$\rightarrow$III at all redshifts.
The highest probability of detecting a GRBII$\rightarrow$III, though, is in halos with a slightly higher SFR, i.e. $\rm \sim 10^{-2.5}~M_{\odot}~yr^{-1}$, as shown in the right column (note that here the distributions are calculated as in Fig.~\ref{fig:Z}, where $Z$ bins are substituted by SFR bins). 
\begin{figure}
  \centering
  \includegraphics[width=1.0\linewidth]{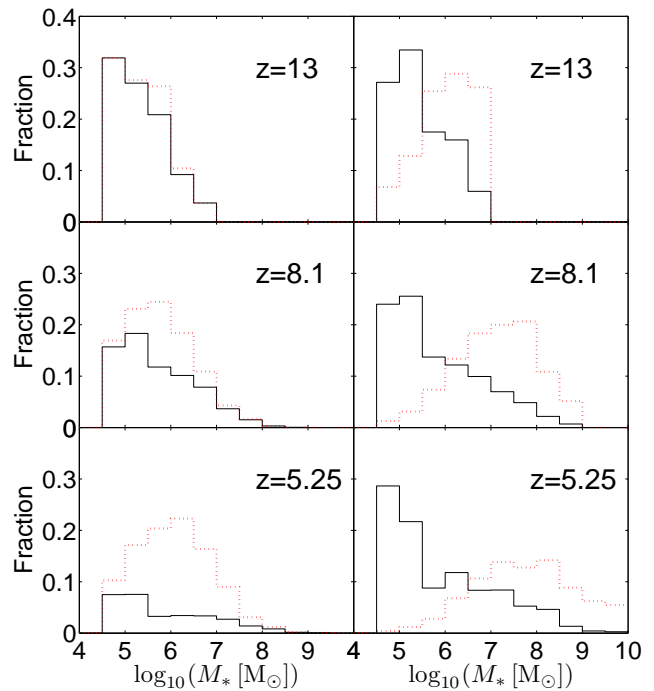}
  \caption{Stellar mass distribution of halos hosting a GRBII$\rightarrow$III (solid black line) and of all PopII/I star forming halos (dotted red line) at redshift $z=13$ (top), 8.1 (middle) and 5.25 (bottom). {\it Left panels:} halos number fraction distribution normalized to the total number of star forming halos. {\it Right panels:} the distributions are ``weighted'' by their respective SFRs and normalized to the total SFR (see text for more details).
  }
  \label{fig:stellarmass}
\end{figure}
\\
Finally, in Figure~\ref{fig:stellarmass} we show the stellar mass distribution of halos hosting a GRBII$\rightarrow$III (solid black line) and of all PopII/I star forming halos (dotted red) at different redshift.
Despite both distributions being very similar in terms of halo number fraction at the highest redshift (left panels)
as $z$ decreases GRBII$\rightarrow$III are still preferentially hosted by halos with $M_\star \rm \sim 10^5~M_\odot$, while the typical host of a GRB has a stellar mass about one or two orders of magnitude larger.
The right panels of the Figure, showing the distributions of the host stellar masses ``weighted'' by their respective halo SFR (note that here the distributions are calculated as in Fig.~\ref{fig:Z} where $Z$ bins are substituted by $M_\star$ bins), confirm that the highest probability of detecting a GRBII$\rightarrow$III is, as above, in structures with $M_\star \rm \sim 10^5~M_\odot$ at all redshifts.

\section{Conclusions}
\label{sec:conclusion}
We use N-body hydrodynamical chemistry cosmological simulations to compute the rate of high-redshift GRBs arising from the death of PopII stars and exploding in an environment previously enriched by PopIII supernovae (GRBII$\rightarrow$III).
Indeed, the detection of these objects can provide an indirect but powerful probe of the existence and the properties of the first stars in the Universe.
\\
The results presented in this work are based on numerical calculations including detailed chemistry evolution, gas cooling according to suited metal-dependent stellar yields (for He, C, N, O, S, Si, Mg, Fe, etc.) and mass-dependent stellar lifetimes for both the pristine PopIII regime and the metal-enriched PopII/I regime \cite[][]{TBDM2007, Maio_etal_2007, Maio_etal_2010}.
Metal pollution proceeds from dense star forming regions to surrounding lower-density environments and is responsible for the enrichment of the Universe already in the first billion years, when early massive stars die and possibly generate GRB events.
Numerical limitations \cite[see a more complete discussion in e.g.][and references therein]{Maio_etal_2010} in addressing such issues might lie in the uncertainties of metal-diffusion modeling (both in SPH and grid-base approaches).
Effects due to the PopII/I IMF slope and features or to the particular critical metallicity adopted in the range $\sim 10^{-6}-10^{-3}\,\rm Z_\odot$ are minor.
\\
On the other hand, a PopIII IMF shifted towards lower masses as recently suggested by some authors \cite[][]{2007ApJ...667L.117Y,2010MNRAS.405..177S,2011Sci...331.1040C,2014ApJ...785...73S} is expected to have a larger impact on our results. In this case, in fact, the metal yields from PopIII stars are expected to be more similar to those from PopII/I stars and thus it would be increasingly difficult (if at all possible) to discriminate gas enriched by primordial stars based on metal abundances.
We explicitly checked this issue by re-running the same box with a PopIII IMF in the $\rm 0.1-100M_{\odot}$ mass range and we 
found results consistent with analogous studies based on DLA observations and employing low-mass pristine IMFs \cite[e.g.][]{2014ApJ...787...64K}.
More specifically, we found that while assuming a SN mass range of 8-40$\mathrm{M}_{\odot}$ it would be extremely difficult to distinguish GRBII$\rightarrow$III because the  metal yields of PopIII stars are too similar to those of PopII/I stars, including stars as massive as 100$\mathrm{M}_{\odot}$ would induce extremely high Oxygen and Carbon yields \cite[][]{2010ApJ...724..341H}. In the latter case, GRBII$\rightarrow$III could be identified e.g. using the condition [Si/O]<-0.6, [S/O]<-0.6 and [C/O]>-0.4. 
\\
Possibly different metal yields as predicted by different stellar structure models could slightly shift the resulting abundance ratios according to the specific input yield tables, but no major changes are expected, mostly at low metallicities \cite[][]{Cescutti2009}.
Resolution effects are generally unavoidable, but they are likely to play a minor role for the abundance ratios explored in this work and for the star formation regime transition, since these processes are predominantly determined by stellar evolution timescales, which are constrained relatively well.
The cosmology adopted is $\Lambda$CDM, but changes in the background cosmological scenario and the introduction of high-order non-linear effects would not affect dramatically the baryonic properties within bound structures \cite[see e.g.][]{Maio2006, Maio2011stream, Maio2012nonG, MaioIannuzzi2011, MaioKhochfar2012, deSouza2013, MaioViel2014arXiv}.
Despite that, some particular dark-matter models might cause a delay in early structure formation \cite[as is the case for warm dark matter;][]{deSouza2013, MaioViel2014arXiv}.
\\
We have found that the rate of GRBII$\rightarrow$III rapidly decreases with cosmic time from $\sim 50$\% at $z=16$ down to less than 1\% at $z<5$. We expect that one tenth of the GRBs exploding at $z=10$ should report the metal signature of the enrichment by a PopIII supernova explosion. By convolving the intrinsic rate of these events with the GRB luminosity function, we found that $\sim 0.06$ GRBII$\rightarrow$III yr$^{-1}$ sr$^{-1}$ should be sufficiently bright to trigger {\tt Swift/BAT}.
This means that $\sim 0.8$ GRBII$\rightarrow$III should be present in the entire {\tt Swift} database.
\\
The identification of such events can be obtained by looking at peculiar metal abundance ratios.
We showed that [C/O] and [Si/O] alone could be enough to distinguish GRBII$\rightarrow$III from other GRB populations.
However, in practice, the detection of more elements (such as S and Fe) could help observers in further confirming the nature of the source.
While we have presented all our results with respect to abundance ratios normalized to oxygen, because O is not very sensitive to theoretical yield uncertainties, it is possible to get particularly interesting hints by referring metal abundances to Si, as well.
In this case, PopII star forming particles enriched by PopIII stars can be identified as those with $\rm [C/Si]<-0.5$, $\rm [O/Si]<0$ and $\rm [Fe/Si]<-0.4$, where the first two conditions exclude gas enriched by SNII, while the last one excludes pollution from SNIa.
AGB stars produce no Si and thus they are excluded in any case from such analysis.
We find that the above conditions give results very similar to those obtained using our reference criteria in the entire redshift range.
\\
It should be noticed that it is hard to get accurate Fe abundances from observations at high redshift, while C, Si and O are usually easier to probe \cite[][]{2013arXiv1312.5631C,2006Natur.440..184K}.
In addition, Fe yields might be overestimated of up to a factor of $\sim 2$ \cite[][]{Cescutti2009}.
Thus, Fe is not the best suited element to identify GRBII$\rightarrow$IIIs.
\\
We also explored the properties of galaxies hosting GRBII$\rightarrow$III in our simulations. 
These are very similar to those that have a high probability to host a GBRIII \cite[][]{2013MNRAS.429.2718S}, i.e.
GRBII$\rightarrow$III are typically found in galaxies with stellar mass ($\rm \sim 10^{4.5} - 10^{7}M_{\odot}$), SFR ($\rm \sim 10^{-3} - 0.1M_{\odot}yr^{-1}$) and metallicity ($\sim 10^{-4} - 10^{-2}Z_{\odot}$) lower than those of galaxies hosting a GRBII.
It is worth to note that galaxies with $\log_{10}\left(Z/Z_\odot\right) < -2.8$ most probably host GRBII$\rightarrow$III and GRBIII.
A GRB with measured metallicity below this $Z$ limit should be considered as a strong candidate for being originated in a PopIII enriched environment (or by a PopIII progenitor).
\\
From an observational point of view, GRBII$\rightarrow$III signatures can be directly investigated for the currently known high-$z$ GRBs.
At $z\gsim 6$ metal absorption lines in the optical-NIR spectrum have been detected and metal abundance ratios measured only for two bursts, namely GRB~050904 at $z=6.3$ \cite[][]{2006Natur.440..184K} and GRB~130606A at $z=5.91$ \cite[][]{2013arXiv1312.5631C}.
\\
The observed abundances for GRB~050904 are [C/H]=-2.4, [O/H]=-2.3, [Si/H]=-2.6 and [S/H]=-1.0 translating into [C/O]=-0.1, [Si/O]=-0.3 and [S/O]=0.7.
On this basis, we can exclude GRB~050904 to be a GRBII$\rightarrow$III.
Recent studies by \cite{2013MNRAS.428.3590T} have reported an updated abundance for S of [S/H]=$-1.6\pm 0.3$ and a corresponding metallicity of $\rm log(Z/Z_{\odot})=-1.6 \pm 0.1$, instead of  $\rm log(Z/Z_{\odot})=-1.3\pm 0.3$ \cite[][]{2006Natur.440..184K} as inferred from Si absorption lines of the afterglow spectrum. However, such variations do not affect our conclusion on GRB~050904.
\\
For GRB~130606A \citet{2013arXiv1312.5631C} reported a 3$\sigma$ upper limit of $\rm [S/H]<-0.82$ and determined the lower limits of the oxygen and silicon abundance as $\rm [Si/H]>-1.80$ and $\rm [O/H]>-2.06$, and metallicity $Z$ as $\sim (1/60 - 1/7)\,{\rm Z}_{\odot}$.
Results from \citet[][]{Chornock2013} give $\rm [Si/H]>-1.7$ and $\rm [S/H]<-0.5$, while \citet[][]{2014arXiv1409.4804H} find $\rm [C/H]>-1.29$, $\rm [O/H]>-1.88$, $\rm [Si/H]=-1.33\pm0.08$, $\rm [S/H]<-0.63$ and $\rm [Fe/H]=-2.12\pm0.08$.
Using these information, it is possible to estimate corresponding limits of $\rm [S/O]<1.24$, $\rm [Si/O]<0.55$, and $\rm [Fe/O]<-0.24$.
These are too weak, though, to probe the nature of the environment in which the burst explodes.
We also note that the lower limits on the total metallicity based on Si and O \cite[][]{Chornock2013} suggest a moderately enriched environment where GRBII are likely to dominate.
However, the possibility that GRB~130606A is associated to a PopIII enriched environment can not be completely excluded, although a comparison with the metallicity expected for a galaxy hosting a GRBII$\rightarrow$III seems to indicate otherwise.
\\
Another possible candidate at $z=4.7$ could be GRB~100219A, for which \cite{2013MNRAS.428.3590T} reported the following measurements: $\rm [C/H]=-2.0\pm 0.2$, $\rm [O/H]=-0.9 \pm 0.5$, $\rm [Si/H]=-1.4 \pm 0.3$,  $\rm [S/H]=-1.1\pm 0.2$ and $\rm [Fe/H]=-1.9 \pm 0.2$. Metallicity is about $\rm ~0.1\,Z_{\odot}$ and is derived from the abundance of sulphur. From such values one can get $\rm [C/O]=-1.1 \pm 0.7$, $\rm [Si/O]=-0.5 \pm 0.8$, $\rm [S/O]=-0.2 \pm 0.7$ and $\rm [Fe/O]=-1 \pm 0.7$. The precision at $1\sigma$-level is not enough to draw robust conclusions, though.
\\
We also note that preliminary analyses are available for GRB~111008A at $z=5.0$ \cite[][]{2014ApJ...785..150S}, which is the highest-redshift GRB with precise measurements of metallicity, column densities and iron, nickel, silicon and carbon fine-structure emissions.
Reported abundance ratios within 2$\sigma$-error bars are:
$\rm [Si/H]> -1.97 $,
$\rm [S/H] =-1.70 \pm 0.10$,
$\rm [Cr/H]=-1.76 \pm 0.11 $,
$\rm [Mn/H]=-2.01 \pm 0.10 $,
$\rm [Fe/H]=-1.74 \pm 0.08 $,
$\rm [Ni/H]=-1.64 \pm 0.19 $,
$\rm [Zn/H]=-1.58 \pm 0.21 $.
The authors do not give carbon and oxygen values, but they note that [Fe/H], [Ni/H] and [Si/H] measurements suggest metallicities around $\sim 1-6$ per cent solar, well above the $10^{-2.8}Z_\odot$ limit for GRBIII and GRBII$\rightarrow$III.
Morevover, the inferred reddening ($\rm A_V =0.11\pm 0.04 \, mag$ ) of the afterglow indicates that the dust-to-metal ratio of the host galaxy is $0.57 \pm 0.26 $, similar to typical values of the Local Group.
Hence, the GRB environment should be strongly dominated by PopII/I enrichment, while  it is unlikely that it has been pre-enriched by massive PopIII stars.
\\
Finally, the metallicity inferred from X-ray measures at higher $z$ \cite[][]{2011MNRAS.410.1611C,2013MNRAS.431.3159S}
suggests values above a few percent solar. These  might not be reliable though, as the X-ray value of the H column density they are based on is likely affected by intervening metals along the line-of-sight \cite[][]{2012MNRAS.421.1697C,2015arXiv150100913C}.
With future X-ray facilities like the {\it Athena} satellite \cite[][]{2013arXiv1306.2307N}\footnote{http://www.the-athena-x-ray-observatory.eu/}
it will become possible  to directly measure the abundance patterns for a variety of ions (e.g. S, Si, Fe) superimposed on bright X-ray afterglows. This, in principle, will allow to discriminate between different nucleo-synthesis sources 
\cite[][]{2013arXiv1306.2336J} and identify GRBII$\rightarrow$IIIs.

\section*{Acknowledgements}
We acknowledge the anonymous referee for careful reading and useful discussions that improved the presentation of the results. U.~M. has received funding from a Marie Curie fellowship by the European Union Seventh Framework Programme (FP7/2007-2013) under grant agreement n. 267251. He also acknoweldges financial support during the finaliazion of this work through a MTG/2015 travel grant of the Marie Curie Alumni Association. We used the tools offered by the NASA Astrophysics Data Systems and by the JSTOR archive for bibliographic research.

\bibliography{ref}

\end{document}